\documentclass[twocolumn,
 reprint,
 amsmath,amssymb,
 aps,
 floatfix,
superscriptaddress
]{revtex4-1}

\usepackage{graphicx}
\usepackage{dcolumn}
\usepackage{bm}
\usepackage[subpreambles=true]{standalone}

\usepackage{array}
\usepackage{float}
\usepackage{calc}
\usepackage{textcomp}
\usepackage{mathtools}
\usepackage{multirow}
\usepackage{amsmath}
\usepackage{amssymb}
\usepackage{graphicx}
\DeclarePairedDelimiter{\ceil}{\lceil}{\rceil}
\usepackage{braket}
\usepackage{qcircuit}
\usepackage[section]{placeins}
\usepackage{flafter}

\begin{document}

\title{Accelerated Variational Quantum Eigensolver}

\author{Daochen Wang}\thanks{wdaochen@gmail.com}
\author{Oscar Higgott}
\author{Stephen Brierley}
\affiliation{
Riverlane, 3 Charles Babbage Road, Cambridge CB3 0GT, United Kingdom
}

\begin{abstract}
The problem of finding the ground state energy of a Hamiltonian
using a quantum computer is currently solved using either the quantum phase estimation (QPE) or variational quantum eigensolver (VQE) algorithms.
For precision $\epsilon$, QPE requires $O(1)$ repetitions of circuits
with depth $O(1/\epsilon)$, whereas each expectation estimation subroutine
within VQE requires $O(1/\epsilon^{2})$ samples from circuits with depth $O(1)$. We propose a generalised VQE algorithm that interpolates between these two regimes via a free parameter $\alpha\in[0,1]$ which can exploit quantum coherence over a circuit depth of  $O(1/\epsilon^{\alpha})$ to reduce the number of samples to $O(1/\epsilon^{2(1-\alpha)})$. Along the way, we  give a new routine for expectation estimation under limited quantum resources that is of independent interest.
\end{abstract}
\maketitle

\section{\label{sec: Introduction} Introduction}

One of the most compelling uses of a quantum computer is to find approximate
solutions to the Schr\"odinger equation. Such ab initio or first-principles
calculations form an important part of the computational chemistry tool-kit and are used to understand features of large molecules
such as the active site of an enzyme in a chemical reaction or are
coupled with molecular mechanics to guide the design of better drugs.

Broadly speaking, there are two approaches to ab initio chemistry
calculations on a quantum computer: one uses the quantum phase estimation
algorithm (QPE) as envisaged by Lloyd~\cite{Lloyd1073} and Aspuru-Guzik et al.~\cite{Aspuru-Guzik2005}, the other uses the variational principle, as exemplified by the variational quantum eigenvalue solver (VQE)~\cite{Peruzzo2014}. Given a fault-tolerant device, QPE can reasonably be
expected to compute energy levels of chemical species as large as the iron
molybdenum cofactor ($\text{FeMoco}$) to chemical accuracy~\cite{Reiher2017}, essential to
understanding biological nitrogen fixation by nitrogenase~\cite{Reiher2017,Hoffman2014}. That QPE \textit{may} provide a quantum-over-classical advantage can be rationalised by the exponential cost involved in naively simulating quantum gates on $n$ qubits by matrix multiplication. One main reason that QPE requires fault tolerance is that the required coherent circuit depth, $D$, scales inversely in the precision $\epsilon$. This means $D=O(1/\epsilon)$ scales exponentially in the number of bits of precision.

The VQE algorithm can also estimate the ground state energy of a chemical Hamiltonian
but does so using a quantum expectation estimation subroutine together
with a classical optimiser. In contrast to QPE, VQE is designed to be run on near-term noisy devices with low coherence time~\cite{Peruzzo2014,McClean2016,OMalley2016}. While VQE may also provide a quantum-over-classical advantage via the same rationalisation as QPE, it suffers from requiring a large number of samples $N=O(1/\epsilon^{2})$ during each expectation estimation subroutine leading to fears that
its run time will quickly become unfeasible~\cite{Wecker2015}. 

We propose a generalised VQE algorithm, we call $\alpha$-VQE, capable of exploiting all available coherence time of the quantum computer to up-to-exponentially reduce the number of samples required for a given precision.  The $\alpha$ refers to a free parameter $\alpha\in[0,1]$ we introduce, such that for all values of $\alpha >0$, $\alpha$-VQE out-performs VQE in terms of the number of samples and has total runtime, $O(N \times D)$, reduced by a factor $O(1/\epsilon^\alpha)$. Moreover, compared to QPE, $\alpha$-VQE has a lower maximum circuit depth for all  $\alpha <1$. At the two extremes, $\alpha =0$ and $\alpha =1$, $\alpha$-VQE recovers the scaling of VQE and QPE respectively.

The $T_{1}$ and $T_{2}$ coherence times of the quantum computer essentially define a maximum circuit depth, $D_{\text{max}}$, that can be run with a low expected number of errors~\footnote{One could alternatively bound the circuit area or total number of quantum gates. We use circuit depth for simplicity.}. By choosing an $\alpha\in[0,1]$ such that
the maximum coherent circuit depth $D(\alpha)=O(1/\epsilon^{\alpha})$ of the expectation
estimation subroutine in $\alpha$-VQE equals
$D_{\text{max}}$,
we show that the expected number of measurements $N$
required can be reduced to $N=f(\epsilon, \alpha)$, where:
\begin{equation}\label{eq:f_epsilon_alpha}
f(\epsilon,\alpha)=\begin{cases}
\frac{2}{1-\alpha}(\frac{1}{\epsilon^{2(1-\alpha)}}-1) & \text{if }\alpha\in[0,1)\\[5pt]
4\,\text{log}(\frac{1}{\epsilon}) & \text{if }\alpha=1
\end{cases}.
\end{equation}

Note that $f(\epsilon,0)=O(1/\epsilon^{2})$ is proportional to the number
of measurements taken in VQE, whereas $f(\epsilon,1)=O(\text{log(1/\ensuremath{\epsilon})})$
is the number of measurements taken in iterative QPE up to further $\text{log}$
factors. 

Our paper is organised as follows. In Sec.~\ref{sec: Generalising VQE}, we generalise VQE to $\alpha$-VQE
by replacing its expectation estimation subroutine with a tunable
version of QPE we name $\alpha$-QPE. This is set out in three steps. In Sec.~\ref{subsec:Tunable Bayesian QPE}, we introduce $\alpha\in[0,1]$ into a Bayesian QPE~\cite{Wiebe2016}
to yield $\alpha$-QPE. Then in Sec.~\ref{subsec:Casting expectation estimation},
we describe how to replace the expectation estimation subroutine
within VQE by $\alpha$-QPE by modifying a result of Knill et al.~\cite{Knill2007}. We end with a schematic illustration of $\alpha$-VQE in Sec.~\ref{subsec:Generalised alpha VQE}. In Sec.~\ref{sec: Resource Comparisons}, we explain how $\alpha$-VQE accelerates VQE.

\section{\label{sec: Generalising VQE} Generalising VQE to $\alpha$-VQE}

The standard VQE algorithm is inspired by the use of variational ansatz
wave-functions $\Ket{\psi(\lambda)}$, depending on a real vector parameter $\lambda$,
in classical quantum chemistry. The ground state energy of a Hamiltonian
$H$ is found by using a hybrid quantum-classical computer to calculate the energy $E(\lambda)$ 
of the system in the state $\Ket{\psi(\lambda)}$, and
a classical optimiser to minimise $E(\lambda)$ over $\lambda$.

The idea is to first write $H$ as the finite sum $H=\sum a_{i}P_{i}$
where $a_{i}$ are real coefficients and $P_{i}$ are a tensor product of 
Pauli matrices.
The number of summed terms is typically polynomial in the system size, as is the case for the electronic Hamiltonian of quantum chemistry. Then for a given (normalised) $\Ket{\psi(\lambda)}$ we estimate the energy:
\begin{equation}\label{weighted sum}
    E(\lambda)\equiv\Bra{\psi(\lambda)}H\ket{\psi(\lambda)}=\sum_{i} a_{i}\Bra{\psi(\lambda)}P_{i}\Ket{\psi(\lambda)},
\end{equation}
using a quantum computer for the individual expectation values and a classical computer for the weighted sum. Finally a classical optimiser is used to optimise the function $E(\lambda)$ with respect to $\lambda$ by controlling a preparation circuit $R(\lambda):\Ket{0}\mapsto\Ket{\psi(\lambda)}$
where $\Ket{0}$ is some fixed starting state. The variational principle justifies the entire VQE procedure: writing $E_{\text{min}}$ for
the ground state eigenvalue of $H$, we have that $E(\lambda)\geq E_{\text{min}}$
with equality if and only if $\Ket{\psi(\lambda)}$ is the ground
state.

Each expectation  $\Bra{\psi(\lambda)}P_{i}\Ket{\psi(\lambda)}$ is directly estimated using statistical sampling~\cite{Romero2017}. The circuit used has extra depth $D=O(1)$ beyond preparing $\Ket{\psi(\lambda)}$ and is repeated $N=O(1/\epsilon^{2})$ times to attain precision within $\epsilon$ of the expectation. Henceforth, we refer to this $N,D$ scaling with $\epsilon$ as the statistical sampling regime.

\subsection{Tunable Bayesian QPE ($\alpha$-QPE) \label{subsec:Tunable Bayesian QPE}}

Since the introduction by Kitaev~\cite{Kitaev} of a type of iterative
QPE involving a single work qubit and an increasing number of controlled
unitaries following each measurement, the term QPE itself has become
associated with algorithms of this particular type. It is characteristic
of Kitaev-type algorithms that for precision $\epsilon$, the number
of measurements $N=\tilde{O}(\log(1/\epsilon))$ and maximum coherent depth $D=\tilde{O}(1/\epsilon)$, where the tilde means we neglect further
$\text{log}$ factors. Henceforth,
we refer to this $N,D$ scaling with $\epsilon$ as the phase estimation regime and QPE as phase estimation in this regime.

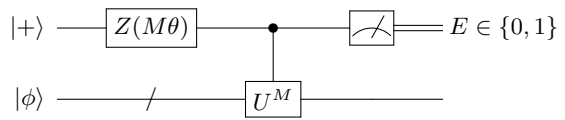
\begin{figure}[H]
\centering 
\mbox{ 
\Qcircuit @C=2em @R=1.4em { \lstick{\ket{+}}  & \gate{Z(M\theta)} & \ctrl{1}  & \meter  & \cw & {\quad E\in \{0,1\}}\\ \lstick{\ket{\phi}}& {/} \qw & \gate{U^M} & \qw & \qw\\}}
\centering{}\caption{{\scriptsize{}Circuit for Kitaev's Phase Estimation and Rejection Filtering Phase Estimation (RFPE). Here, $\Ket{\phi}$ is an
eigenstate of $U$ with eigenphase $\phi$,
$\Ket{+}$ is the $+1$ eigenstate of $X$, $Z(M\theta)\coloneqq\text{diag}(1,e^{-iM\theta})$, and measurement is performed
in the $X$ basis.}}
\label{RFPE circuit}
\end{figure}

For a given eigenvector $\Ket{\phi}$ of a unitary operator
$U$ such that $U\Ket{\phi}=e^{i\phi}\Ket{\phi},\phi\in[-\pi,\pi)$, Kitaev's QPE algorithm uses the circuit in Fig.~\ref{RFPE circuit} with two
settings of $M\theta\in\{0,-\pi/2\}$. For each setting, $N=\tilde{O}(\text{log}(1/\epsilon))$ measurements are taken with $M = 2^{m-1}, 2^{m-2},...,1$ in that order to estimate $\phi$ to precision $\epsilon\equiv2^{-m}$. In Kitaev's algorithm, ``precision $\epsilon$'' means ``within error $\epsilon$ above a constant level of probability''.  The coherent circuit depth $D$ required is therefore:
\begin{equation}\label{eq:kitaev_qpe_depth}
D=\tilde{O} \left( \sum_{j=0}^{m-1}{2^{j}} \right) = \tilde{O} \left( 2^{m} \right)=\tilde{O} \left(1/\epsilon \right).
\end{equation}

This accounting associates to $U^{2^{j}}$ a circuit depth of $O(2^{j})$. For generic $U=\text{exp}(-iHt)$, any better accounting is prohibited by the ``no-fast-forwarding" theorem~\cite{Berry2007}. We do not consider special $U$ such that $U^{2^{j}}$ has better accounting (e.g. modular multiplication in Shor's algorithm~\cite{Nielsen2010}).

Under the framework of Kitaev's QPE, Wiebe and Granade~\cite{Wiebe2016, Wiebe2015} introduced a Bayesian QPE named Rejection Filtering Phase Estimation (RFPE) which we now modify to yield different sets of circuit and measurement sequences that can provide the same precision $\epsilon$ with different $(N,D)$ trade-offs. It is these sets that shall be parametrised by the $\alpha\in[0,1]$. The circuit for RFPE is given in Fig.~\ref{RFPE circuit} and the following presentation of RFPE and our modification is broadly self-contained.

To begin, a prior probability distribution $P(\phi)$ of $\phi$ is taken to be normal $\mathcal{N}(\mu,\sigma^2)$ (some
justification is given in Ref.~\cite{Ferrie2013} which empirically found
that the posterior of a uniform prior converges rapidly to normal).
From the RFPE circuit in Fig.~\ref{RFPE circuit}, we deduce the probability of measuring $E\in\left\{ 0,1\right\}$
is: 
\begin{equation}\label{eq:rfpe_meas_prob}
P(E|\phi;M,\theta)=\frac{1+(-1)^{E}\text{cos}(M(\phi-\theta))}{2},
\end{equation}
which enters the posterior by the Bayesian update rule:
\begin{equation}\label{eq:bayesian_update}
P(\phi|E;M,\theta)\propto P(E|\phi;M,\theta)P(\phi).
\end{equation}

We do not need to know the constant of proportionality to sample from
this posterior after measuring $E$, and the word ``rejection''
in RFPE refers to the rejection sampling method used. After obtaining
a number $s$ of samples, we approximate the posterior again by a normal
with mean and standard deviation equal to that of our samples (again justified as when taking initial prior to be normal). The choice
of $s$ is important and $s$ can be regarded as a particle filter
number, hence the word ``filter'' in RFPE~\cite{Wiebe2015}. We constrain posteriors to be normal because normal distributions can be efficiently sampled.

The effectiveness of RFPE's iterative update procedure just described depends on controllable parameters $(M,\theta)$. A natural measure of effectiveness is the expected posterior variance, i.e. the ``Bayes risk''. To minimise the Bayes risk, Ref.~\cite{Wiebe2016} chooses $M=\ceil{1.25/\sigma}$ at the start of each iteration. However, the main problem is that $M$ can quickly become large, making the depth of $U^{M}$ exceed $D_{\text{max}}$. Ref.~\cite{Wiebe2015} addresses this problem by imposing an upper bound on $M$ and we refer to this approach as RFPE-with-restarts. 

Here, we propose another approach that chooses: 
\begin{equation}\label{eq:m_and_theta}
(M,\theta)= \left( \frac{1}{\sigma^\alpha}, \ \mu-\sigma \right),
\end{equation}
where $\alpha\in[0,1]$ is a free parameter we impose. 
Moreover, we propose a new preparation of eigenstate $\Ket{\phi}$ at each iteration, discarding that used in the previous iteration. This ability to readily prepare an eigenstate is highly atypical but can be achieved within the VQE framework (see Sec.~\ref{subsec:Casting expectation estimation}). 
We name the resulting, modified RFPE algorithm $\alpha$-QPE. In Proposition~1 below, we give the main performance result about $\alpha$-QPE. We defer its derivation to the Supplementary Material~\cite{SuppMat}. Unlike in Kitaev's algorithm, we henceforth let ``precision $\epsilon$'' mean an expected posterior standard deviation of $\epsilon$~\footnote{An actual standard deviation of $\epsilon$ on an unbiased posterior mean implies ``precision $\epsilon$'' in Kitaev's sense by Markov's inequality. The converse is not true. In the Supplementary Material~\cite{SuppMat}, we numerically verify that our new definition of $\epsilon$ well approximates the true error.}.

\vspace{2 mm}
\textbf{Proposition~1.}---(Measurement--depth trade-off). For precision $\epsilon$, $\alpha$-QPE requires: $N = f(\epsilon,\alpha)$ measurements and $D =O(1/\epsilon^{\alpha})$ coherent depth, where the function $f$ is defined in Eqn.~\ref{eq:f_epsilon_alpha}.
\vspace{2 mm}

We now address the essential question of how to choose $\alpha$ when practically constrained to circuits with bounded depth $D\in[1,D_{\text{max}}]$ for some $D_\text{max}$. For simplicity, we assume $D=1/\epsilon^\alpha$. Optimally choosing $\alpha$ amounts to minimising the number of measurements $N$ to achieve a fixed precision $\epsilon\in(0,1)$. Then, because
$N=f(\epsilon,\alpha)$ is a decreasing function
of $\alpha$, the least $N$ is attained at the maximal $\alpha=\alpha_{\text{max}}\coloneqq\text{min}\left\{\frac{\text{log}(D_{\text{max}})}{\text{log}(1/\epsilon)},1\right\}$,
giving $N_{\text{min}} = f(\epsilon, \alpha_{\text{max}})$ which equals:

\begin{equation}\label{eq:alpha_qpe_n_min}
\begin{cases}
\frac{2}{1-\text{log}(D_{\text{max}})/\text{log}(1/\epsilon)}((\frac{1}{\epsilon D_{\text{max}}})^{2}-1) & \text{if }D_{\text{max}}<\frac{1}{\epsilon}\\[5pt]
4\,\text{log}(\frac{1}{\epsilon}) & \text{if }D_{\text{max}}\geq\frac{1}{\epsilon}
\end{cases}.
\end{equation}

The important point here is the inverse quadratic scaling with $D_{\text{max}}$ if $D_{\text{max}}<1/\epsilon$: through $\alpha$ we can access and exploit $D_{\text{max}}$ to significantly reduce the number
of iterations. In the Supplementary Material~\cite{SuppMat}, we deduce from our above analysis that RFPE is at least as efficient as Eqn.~\ref{eq:alpha_qpe_n_min}.

\subsection{Casting expectation estimation as $\alpha$-QPE \label{subsec:Casting expectation estimation} }

Given a Pauli operator $P$, a preparation circuit $R(\lambda)\equiv R:\Ket{0}\mapsto\Ket{\psi(\lambda)}\equiv\Ket{\psi}$,
and a projector $\Pi\coloneqq I-2\Ket{0}\Bra{0}$, we paraphrase from Knill
et al.~\cite{Knill2007} the following Proposition 2 relevant to us. 

\vspace{2 mm}
\textbf{Proposition~2.}---(Amplitude estimation).
The operator $U:=U_{0}U_{1}$, with $U_{0}=(R\Pi R^{\dagger}),U_{1}=(PR\Pi R^{\dagger}P^{\dagger})$,
is a rotation by an angle $\phi=2\,\text{arccos}(\left|\Bra{\psi}P\Ket{\psi}\right|)$
in the plane spanned by $\Ket{\psi}$ and $\Ket{\psi'}\coloneqq P\Ket{\psi}$.
Therefore, the state $\Ket{\psi}$ is an equal superposition of eigenstates $\Ket{\pm \phi}$
of $U$ with eigenvalues $e^{\pm i\phi}$ respectively (i.e. eigenphases $\pm\phi$)
and we can estimate $\left|\Bra{\psi}P\Ket{\psi}\right|=\text{cos}(\pm\phi/2)$
to precision $\epsilon$ by running QPE on $\Ket{\psi}$ to precision
$2\epsilon$. %
\vspace{2 mm}

Note that the VQE framework readily provides $R(\lambda)$ which enables our use of Proposition 2. We now modify Proposition 2 to use $\alpha$-QPE which enables access to the measurement-depth trade-off given in Proposition 1. Since $\alpha$-QPE requires re-preparation of state $\Ket{\pm\phi}$ at each iteration, a complication arises because $\Ket{\psi}$ is in equal superposition of $\Ket{\pm\phi}$. To be able to efficiently collapse $\Ket{\psi}$ into one of $\Ket{\pm\phi}$ with high confidence before each iteration in $\alpha$-QPE, we have to assume that $|A|$ is always bounded away from $0$ and $1$ by a constant $\delta>0$, where $A=\Bra{\psi}P\Ket{\psi}$ (see Ref.~\cite[Parallelizability]{Knill2007}). If we collapse into $\Ket{\phi}$ (with high confidence), we implement $\alpha$-QPE using (powers of) $c\text{-}U$; else if we collapse into $\Ket{-\phi}$, we use $c\text{-}U^{\dagger}$. The depth overhead of state collapse is $O(1/\delta)$. A second complication is that $\phi$ gives $|A|$ but not the sign of $A$. 

These two complications can be simultaneously resolved using a simple two-stage method. In the first stage,  $A$ is roughly estimated by statistical sampling a constant number of times to determine whether $|A|$ satisfies a $\delta$ bound. If so, then proceed with $\alpha$-QPE, else continue with statistical sampling in the second stage. The first stage simultaneously determines the sign of $A$. In the Supplementary Material~\cite{SuppMat}, we present further details of this method.

The overhead in implementing $c\text{-}U=  R (c\text{-} \Pi) R^{\dagger} P R (c\text{-} \Pi) R^{\dagger} P$ is documented as follows. Since $P$ is $n$ tensored Pauli matrices, it can be implemented using $n$ parallel Pauli gates in $O(1)$ depth. The $(n+1)$-qubit controlled sign flip $c\text{-}\Pi$
is equivalent in cost, up to $\sim2n$
single qubit gates with $O(1)$ depth, to an $(n+1)$-bit Toffoli gate, the best known implementation of which requires $6n-6$ CNOT gates~\footnote{In our pre-fault-tolerant setting, the CNOT gate count is the most relevant resource count.}, $\ceil{\frac{n-2}{2}}$ ancillas and $O(\log{n})$ circuit depth~\cite{Maslov2016}. Lastly, we need two $R$ and two $R^{\dagger}\equiv R^{-1}$. Since the depth $C_\text{R}$ of $R$ is $\Omega(n)$ in most applications considered so far~\cite{Babbush2018}, this last overhead may be the most significant. As the total overhead has no $\epsilon$ dependence, it does not affect our analysis in terms of $\epsilon$.
\vspace{-0.2cm}
\subsection{Generalised $\alpha$-VQE \label{subsec:Generalised alpha VQE}}

\begin{figure}[H]
\includegraphics[scale=1]{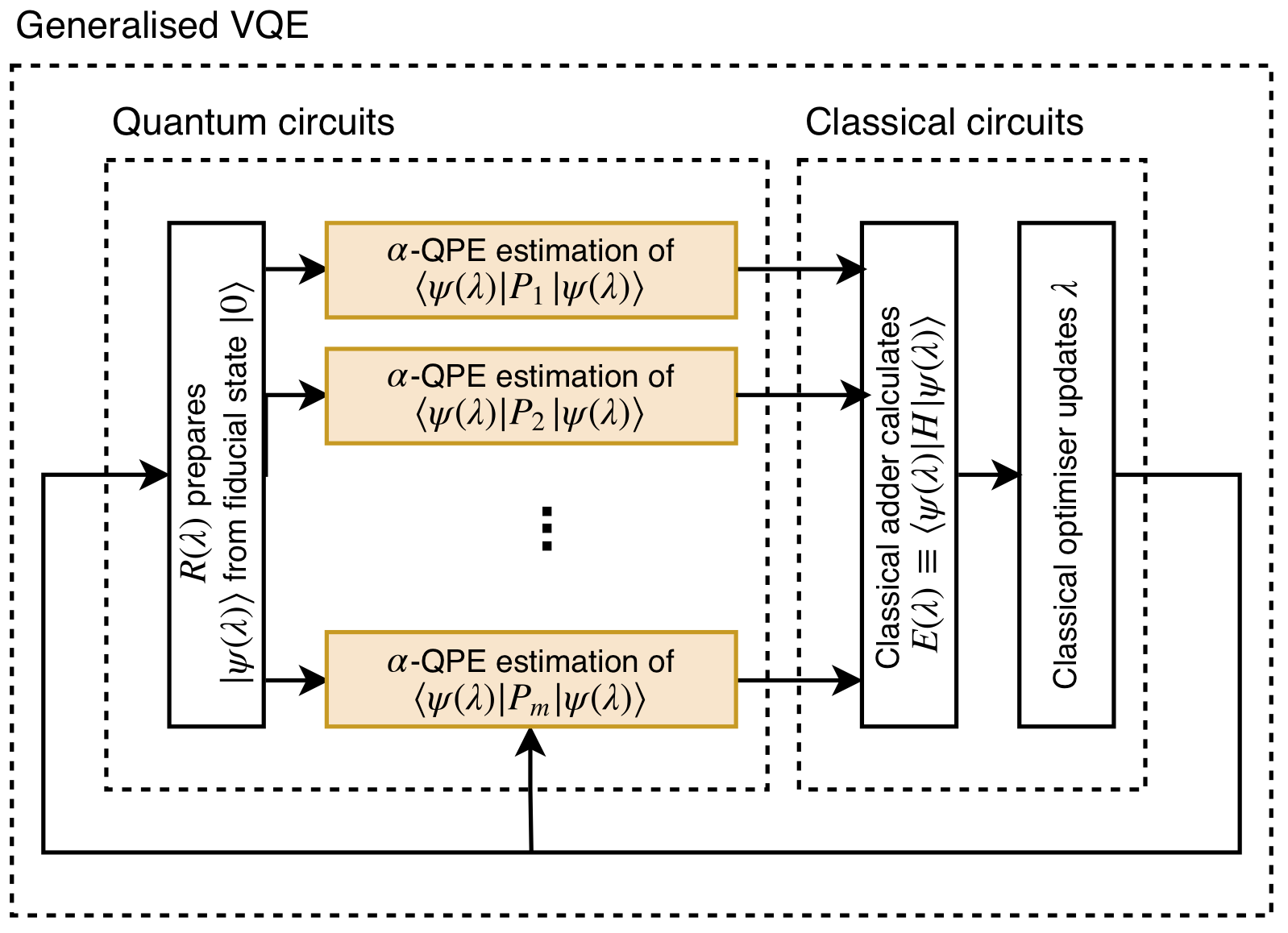}
\centering{}\caption{{\scriptsize{}Schematic of $\alpha$-VQE. Note that $\lambda$
also affects $\alpha$-QPE circuits which involve state preparation $R(\lambda)$ and its inverse.
When $\alpha=0$, we are in the statistical sampling, or standard
VQE, regime. When $\alpha=1$, we are in the phase estimation regime.}}
\label{Generalised VQE}
\end{figure}

We define generalised $\alpha$-VQE by using the result of Sec.~\ref{subsec:Casting expectation estimation}
to replace the method of expectation estimation in VQE by the $\alpha$-QPE developed in Sec.~\ref{subsec:Tunable Bayesian QPE}. Fig.~\ref{Generalised VQE} illustrates the schematic of our generalised VQE.  

The total number of measurements in an entire run of $\alpha$-VQE is of order $f(\epsilon,\alpha)$ multiplied by both the number of summed terms in the Hamiltonian and the number of iterations of the classical optimiser. Writing $C_{R}$ for the depth of $R(\lambda)$, each measurement results from a circuit of depth $O((C_{R}+\log{n})/\epsilon^{\alpha})$. 

Clearly, $\alpha$-VQE still preserves the following three key advantages
of standard VQE because we
only modified the expectation estimation subroutine. First, we can parallelise the expectation estimation of multiple Pauli terms to multiple processors. Second, robustness via self-correction is preserved because $\alpha$-VQE
is still variational~\cite{OMalley2016,McClean2016}. Third, the variational parameter $\lambda$ can be classically stored to enable straightforward re-preparation of $\Ket{\psi(\lambda)}$~\cite{Wecker2015}. 

\section{\label{sec: Resource Comparisons} $\alpha$-VQE as accelarated VQE}

We reiterate that $\alpha$-VQE is useful because it can perform expectation estimation in regimes lying continuously between statistical sampling and phase estimation. Neither extreme is ideal: statistical sampling requires $N=O(1/\epsilon^{2})$
samples whereas phase estimation requires $D=O(1/\epsilon)$ coherence
time. In this manner, these two extremes have been criticised in Ref.~\cite{Paesani2017} and Ref.~\cite{Peruzzo2014,McClean2016} respectively, and compared in Ref.~\cite{Wecker2015}. 

The resources required for one run of expectation estimation within VQE and $\alpha$-VQE (arbitrary $\alpha$,
$\alpha=0$, $\alpha=1$) are compared in Table~\ref{Resource table}.
Neglecting the small overheads to cast expectation estimation as $\alpha$-QPE, we can conclude that our method of expectation estimation is always superior to statistical sampling for $\alpha>0$.

\begin{figure}[H]
\includegraphics[scale=0.5]{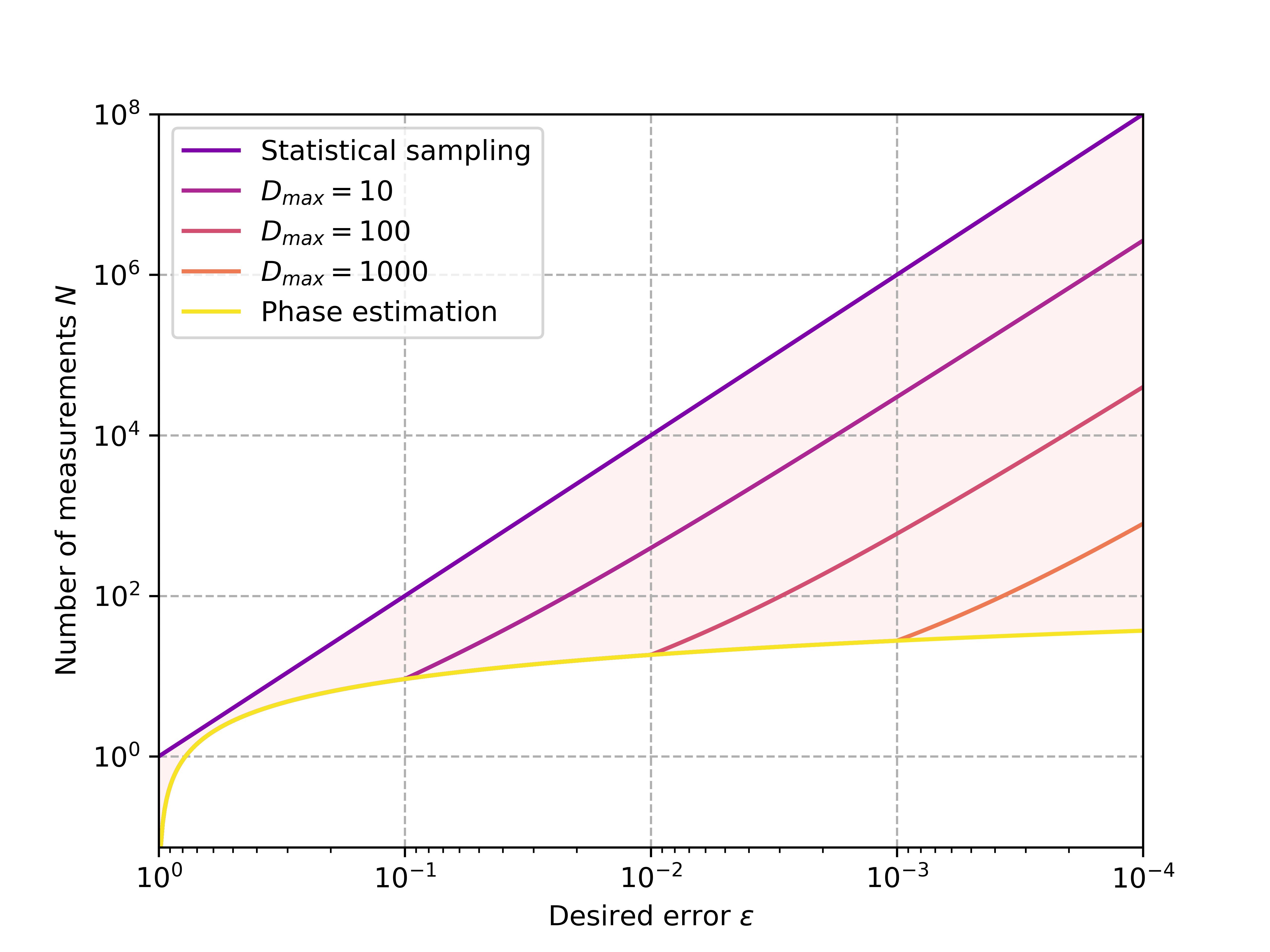}
\centering{}
\caption{{\scriptsize{} Plots of the function in Eqn.~\ref{eq:alpha_qpe_n_min} for different $D_\text{max}$ demonstrate how $\alpha$-VQE accelerates VQE by reducing the number of measurements up-to-exponentially as $D_\text{max}$ increases. Also plotted are the statistical sampling and phase estimation regimes. $\alpha$-VQE unlocks regimes in the shaded region between these two extremes.}}
\label{applicability}
\end{figure}

To use $\alpha>0$, we need sufficiently large $D_\text{max}$. Conversely, given $D_\text{max}$ we can choose an $\alpha$ to maximally exploit it as per our analysis at the end of Sec.~\ref{subsec:Tunable Bayesian QPE}. This provides the mechanism by which $\alpha$-VQE accelerates VQE. The acceleration is quantified by Eqn.~\ref{eq:alpha_qpe_n_min}. We plot Eqn.~\ref{eq:alpha_qpe_n_min} in Fig.~\ref{applicability} to give a concrete sense of our contribution. 

At a more theoretical level, we note that our paper can be viewed outside the VQE context as a study of efficient expectation estimation under restricted circuit depth. Furthermore, Sec.~\ref{subsec:Tunable Bayesian QPE} of our paper can be viewed as a study of phase estimation under restricted circuit depth. Subsequently to our paper, Ref.~\cite{Terhal2018} also studied this latter question, proposing and analysing a time series estimator which learns the phase with similar efficiency as our results. More precisely, their efficiency Eqn.~22 conforms to our Eqn.~\ref{eq:alpha_qpe_n_min} up to log factors.

\section{\label{sec: Acknowledgements} Acknowledgements}
We thank Mark Rowland and Jarrod McClean for insightful discussions.

\begin{table*}
\begin{centering}
\begin{tabular}[b]{c|c|c|c}
\multirow{1}{*}{Algorithm} & Maximum coherent depth & Non-coherent repetitions & Total runtime\tabularnewline
\hline 
VQE & $O(C_{R})$ & $O(\frac{1}{\epsilon^{2}})$ & $O(C_{R}\,\frac{1}{\epsilon^{2}})$\tabularnewline

$0$-VQE & $O(C_{R}+\log{n})$ & $O(\frac{1}{\epsilon^{2}})$ & $O((C_{R}+\log{n})\,\frac{1}{\epsilon^{2}})$\tabularnewline

$1$-VQE  & $O((C_{R}+\log{n})\,\frac{1}{\epsilon})$ & $O(\log\frac{1}{\epsilon})$ & $O((C_{R}+\log{n})\frac{1}{\epsilon})$\tabularnewline

$\alpha$-VQE & $O((C_{R}+\log{n})\,\frac{1}{\epsilon^{\alpha}})$ & $O(f(\epsilon,\alpha))$ & {$O((C_{R}+\log{n})\,\frac{1}{\epsilon^{\alpha}}f(\epsilon,\alpha))$}
\end{tabular}
\par\end{centering}
\caption{{\scriptsize{}Resource comparison of one expectation estimation subroutine within VQE, $0$-VQE, $1$-VQE, $\alpha$-VQE. $\epsilon$ is the precision required for the expected energy, $C_{R}$ is the state preparation depth, and $\alpha\in[0,1]$ is the free
parameter controlling the maximum circuit depth of $\alpha$-QPE.}}
\label{Resource table}
\end{table*}
\FloatBarrier
\bibliographystyle{apsrev4-1}
\bibliography{BibResubmit020319}
\FloatBarrier
\newpage
\appendix
\title{Supplementary Material for\\
``Accelerated Variational Quantum Eigensolver''}
\onecolumngrid
\vspace{-1cm}
\section{Derivation of Proposition 1}\label{subsec:Derivation of Proposition 1}

To analyse RFPE's convergence, we analyse the expected posterior variance $r^{2}$ (i.e. the Bayes
risk) for a normal prior $\phi\sim\mathcal{N}(\mu,\sigma^{2})$. The formula for $r^2$ can be derived from Ref. \cite[Appendix B]{Ferrie2013} as:

\begin{equation}\label{eq:9}
\mathbb{E}_{E}[\mathbb{V}[\phi|M,\theta;\mu,\sigma]]\equiv r^{2}(M,\theta;\mu,\sigma)\equiv r^{2}(M,\theta)\equiv r^{2}=\sigma^{2}(1-\frac{M^{2}\sigma^{2}\text{sin}^{2}(M(\mu-\theta))}{e^{M^{2}\sigma^{2}}-\cos^{2}(M(\mu-\theta))}).
\end{equation}

Note that $r^{2}$ is bounded below by an envelope $s^{2}\coloneqq \sigma^{2}(1-M^{2}\sigma^{2}e^{-M^{2}\sigma^{2}})$. As a function of $M$, $s^2$ has minimiser:
\begin{equation}\label{eq:10}
M_0=\frac{1}{\sigma}.
\end{equation}
But $M_0$ may be far away from the minimiser $M_1$ of $r^{2}$
due to rapid oscillations of $r^{2}$, as a function of $M$, above the envelope $s^2$. Fortunately, the frequency of these oscillations is controlled by $\theta$. This control is exactly the reason why Ref.~\cite{Wiebe2014} introduced $\theta$. Numerical simulations in Ref. \cite[Appendix C]{Wiebe2014} showed that the optimal $\theta\approx\mu\pm\sigma$ can effectively remove oscillations from $r^2$. This aligns $r^2$ with its envelope $s^2$, forcing $M_1$ closer to $M_0$. 

Therefore, it makes sense to choose $(M=1/\sigma,\theta=\mu\pm\sigma)$ if we wish to minimise $r^2(M,\theta)$. However, Ref.~\cite{Wiebe2014} did not give intuition. To gain intuition, we found a simple heuristic argument for why it makes sense to choose $(M\propto 1/\sigma, \theta=\mu\pm\sigma)$ if we wish to minimise $r^2(M,\theta)$. We present our argument in the box below. 

\vspace{5 mm}

\noindent\fbox{\begin{minipage}[c]{1\columnwidth - 2\fboxsep - 2\fboxrule}%
\textbf{Optimal $M,\theta$}
\vspace{1 mm}

We heuristically justify the optimality (in RFPE) of both $\theta\approx\mu\pm\sigma$
and the form $M\propto1/\sigma$ at each iteration using the following simple argument.
Recall that the probability of measuring $E=0$ in the RFPE circuit
is:
\begin{equation}\label{eq:11}
P_{0}=P(0|\phi;M,\theta)=\frac{1+\text{cos}(M(\phi-\theta))}{2}.
\end{equation}

In order to gain maximal information about $\phi$, it is intuitively
obvious that the range of $P_{0}$ has to uniquely and maximally vary
across the domain of uncertainty in $\phi$. The Bayesian RFPE conveniently
gives this domain $\mathcal{D}=(\mu-\sigma,\mu+\sigma)$ of uncertainty
at each iteration. A naive domain on which the range of $\text{cos}$
uniquely and \textit{possibly} maximally varies is $[0,\pi]$. So we would like to control $(M,\theta)$ such that $M(\mathcal{D}-\theta)$
is equal to $[0,\pi]$, i.e.

\begin{equation}\label{eq:12}   
\begin{cases}     M(\mu-\sigma-\theta)=0, \\[5pt]     M(\mu+\sigma-\theta)=\pi.   \end{cases} \end{equation}

This has solution: 
\begin{equation}\label{eq:13}
(M,\theta)=(\frac{\pi/2}{\sigma},\mu-\sigma),
\end{equation}
which is not far from the optimal choice found in
Ref. \cite[Appendix C]{Wiebe2014}. Intuitively, the slight discrepancy
could only be due to $[0,\pi]$ not being the domain on which cosine
(uniquely and) maximally varies. 

\setlength\parindent{24pt}
\end{minipage}}

\vspace{5 mm}

Therefore, we choose $\theta=\mu\pm\sigma$ and trial $M=a/\sigma$ with $a\in\mathbb{R}$
in Eqn.~\ref{eq:9} to give: 

\begin{equation}\label{eq:14}
r^{2}(\frac{a}{\sigma},\mu\pm\sigma)=\sigma^{2}(1-g(a)),
\end{equation}
where $g:\mathbb{R}\rightarrow\mathbb{R}$ is defined by:
\begin{equation}\label{eq:15}
g(x)\coloneqq\frac{x^{2}\text{sin}^{2}(x)}{e^{x^{2}}-\text{cos}^{2}(x)}.
\end{equation}

\begin{figure}
\includegraphics[scale=0.5]{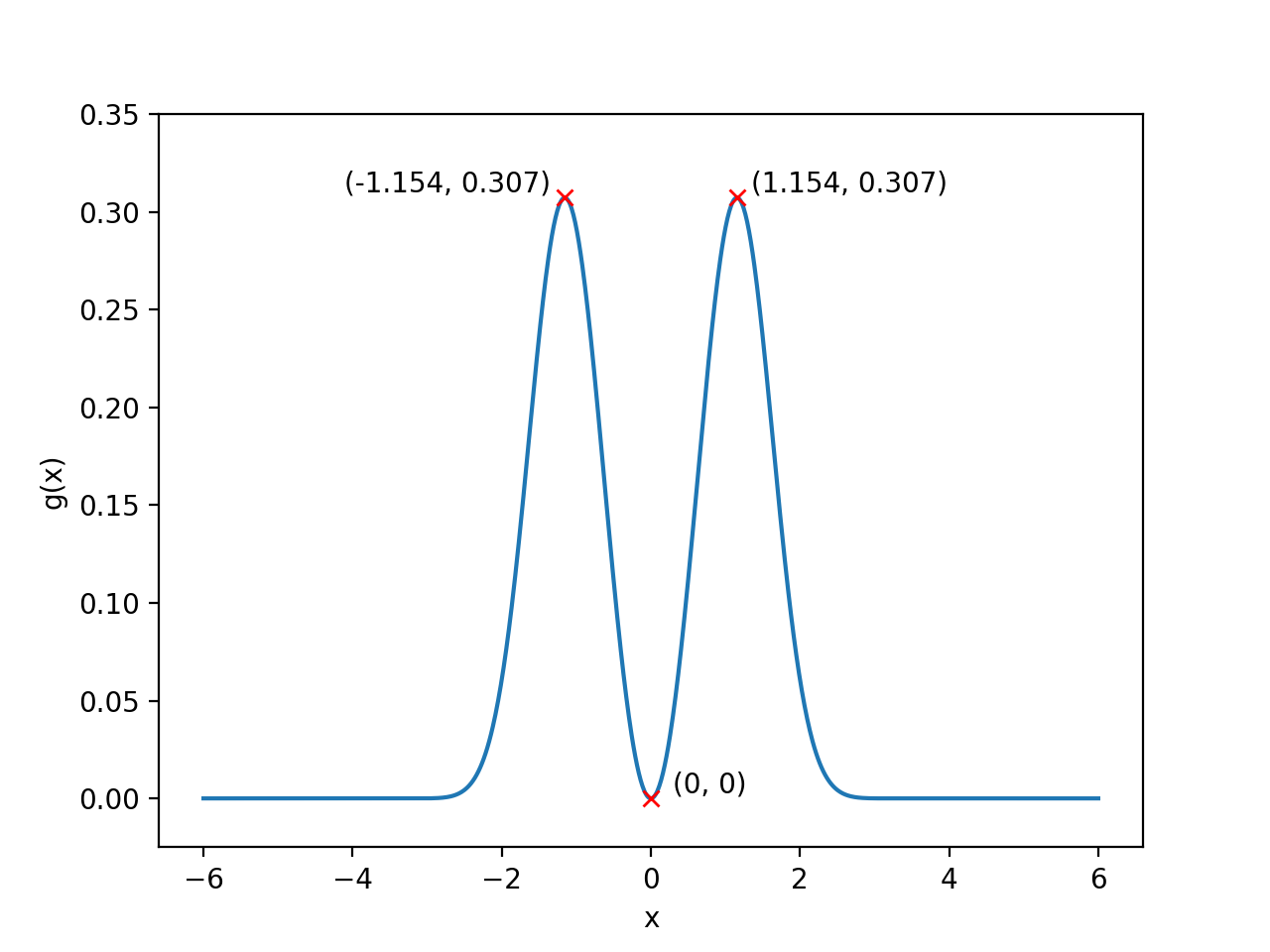}
\centering{}\caption{{\scriptsize{}Plot of $g(x)=\frac{x^{2}\text{sin}^{2}(x)}{e^{x^{2}}-\text{cos}^{2}(x)}$.
$g$ has maxima at $\approx(\pm a_{0},0.307)$ where $a_0\approx 1.154$ and minimum
at $(0,0)$. Near $x=0$, $g(x)=x^{2}/2+O(x^{4})$. }}
\label{g plot}
\end{figure}

We find that $g$ has maximum value $g_{\text{max}}\approx0.307$ at $x=\pm a_0$ where $a_0\approx1.154$, and so $r^{2}$ has
minimum value:
\begin{equation}\label{eq:16}
r_{\text{min}}^{2}=L^{2}\,\sigma^{2},
\end{equation}
where $L^2\approx0.693$. Therefore, after each iteration
of RFPE, we expect the variance to (at least) decrease by a factor
of $L^2$ when $M$ and $\theta$ are chosen optimally~\footnote{Locally optimal $(M,\theta)$ at each
iteration may not be globally optimal over a number of iterations. In fact, $A\approx1.154$ differs from the globally optimal heuristic of $1.25$, but this distinction between local
and global is besides the main point here
and shall not be further discussed.}.

Writing $\sigma_{k}$ for the standard deviation at the $k$-th iteration,
we rewrite Eqn.~\ref{eq:16} as $\mathbb{E}[\sigma_{k}^{2}|\sigma_{k-1}^{2}]=L^{2}\,\sigma_{k-1}^{2}.$
Taking expectation over $\sigma_{k-1}$ gives
$\mathbb{E}[\sigma_{k}^{2}]=L^{2}\,\mathbb{E}[\sigma_{k-1}^{2}]$. Assuming that $\mathbb{V}[\sigma_{k}]=0$ for $k$ large~\footnote{We heuristically justify this and subsequent assumptions or approximations
by good agreement of our final results Eqns.~\ref{al:2}, \ref{eq:28} with
numerical simulations.}, say $k\geq n_0$, we commute squaring with expectation to give
$\mathbb{E}[\sigma_{k}]=L^{(k-k_{0})}\,\mathbb{E}[\sigma_{k_{0}}].$
Writing $r_{k}\coloneqq\mathbb{E}[\sigma_{k}]$ for the expected standard
deviation at the $k$-th iteration gives:
\begin{equation}\label{eq:17}
r_{k}=L^{(k-k_{0})}\,r_{k_{0}},
\end{equation}
so we expect the standard deviation to decrease exponentially with
the number of iterations of RFPE.

Since $r_{k}$ of RFPE decreases exponentially with $k$, the use
of $M\propto1/\sigma_{k}$ at the $k$-th iteration means we expect
$M$ to increase exponentially with $k$. This means that RFPE is
indeed in the phase estimation regime which still has the same problem
of requiring an exponentially long coherence time in the number of
bits of precision required. 

In the following, we address this problem by modifying the dependence
of the $M$ on $\sigma$ at each iteration. We note that a
possible additional restarting strategy in RFPE also addresses this
same problem (see Appendix~\ref{subsec:RFPE-with-restarts}) but for now, RFPE refers to
RFPE \textit{without} restarts. 

Note that RFPE uses $M=O(1/\sigma)$ and is in the phase estimation
regime, but if $M=O(1)$ at each iteration, we expect to recover the
statistical sampling regime. We are led naturally then to consider
$M$ of form:
\begin{equation}\label{eq:18}
M=a(\frac{1}{\sigma})^{\alpha},
\end{equation}
with an introduced $\alpha\in[0,1]$ and some $a=a(\alpha)\in\mathbb{R}$
to facilitate a transition between the two regimes.

We again substitute $\theta=\mu\pm\sigma$, but $M$
as in Eqn.~\ref{eq:18}, into Eqn.~\ref{eq:9}, giving expected posterior variance:
\begin{equation}\label{eq:19}
r^{2}(a(\frac{1}{\sigma})^{\alpha},\mu\pm\sigma)=\sigma^{2}(1-g(b)),
\end{equation}
where $b\coloneqq a\sigma^{(1-\alpha)}$ and $g$ remains defined by Eqn.~\ref{eq:15}.
Ideally, we would like $b=a_{0}$ which gives $a=a_{0}(1/\sigma)^{(1-\alpha)},$
but we need $a$ to be independent of $\sigma$. From the graph of
$g$ (Fig.~\ref{g plot}), we see there is no natural way to
define an optimal $a=a(\alpha)$ except when $\alpha=1$. So we could
simply take $a=a_{0}$ (independent of $\alpha$) but instead
we set $a=1$ for simplicity. 

In the remainder of Appendix~\ref{subsec:Derivation of Proposition 1}, $\alpha\neq1$ ($\alpha=1$
already analysed above) unless stated otherwise and we assume $r_{k}$ converges to zero. This is necessary
for valid Taylor approximations and divisions by $(1-\alpha)$. 

For $\sigma$ small, and so $b$ small, we have:
\begin{equation}\label{eq:20}
g(b)=\frac{b^{2}}{2}+O(b^{4}),
\end{equation}
which we substitute into Eqn.~\ref{eq:19} to give the following upon taking
expectations and using the earlier assumption that $\mathbb{V}[\sigma_{k}]=0$ for $k$ large to commute the expectation:
\begin{equation}\label{eq:21}
r_{k+1}^{2}=r_{k}^{2}(1-\frac{1}{2}(r_{k}^{2})^{1-\alpha}),
\end{equation}
which is similar to a logistic map in $r_{k}^2$. Taking $\text{log}$ gives $\text{\text{log}}(r_{k+1}^2)=\log(r_{k}^2)-\frac{1}{2}r_{k}^{2(1-\alpha)},$ to $O(r_{k}^{4(1-\alpha)})$, which gives, upon writing $l_{k}=\text{log}(r_{k}^2)$:
\begin{equation}\label{eq:23}
l_{k+1}=l_{k}-\frac{1}{2}e^{(1-\alpha)l_{k}}.
\end{equation}
Assuming the existence of a differentiable function $l=l(t)$ with $l(t_{k})=l_{k}$
where $t_{k}\coloneqq nh$, we substitute $l$ into Eqn.~\ref{eq:23} to
obtain:
\begin{equation}\label{eq:24}
\frac{l(t_{k}+h)-l(t_{k})}{h}=\frac{-e^{(1-\alpha)l(t_{k})}}{2h}.
\end{equation}
We further take  $h$  small and assume LHS Eqn.~\ref{eq:24} is well approximated
by a derivative~\footnote{This may be inconsistent with the previous assumption
because it requires $l(t_{k}+h)-l(t_{k})\equiv l_{k+1}-l_{k}=O(h)$
and we assess its consequences in Eqn.~\ref{eq:26}.}. Solving the resulting differential equation under initial condition at $(k_{0},r_{k_{0}})$ gives:
\begin{equation}\label{eq:25}
\text{log}(r_{k})=\text{log}(r_{k_{0}})-\frac{1}{2(1-\alpha)}\text{log}(1+r_{k_{0}}^{2(1-\alpha)}\frac{1-\alpha}{2}\,(k-k_{0})).
\end{equation}
To assess Eqn.~\ref{eq:25} with respect to the recurrence Eqn.~\ref{eq:23} it
intended to solve, we substitute it back to give:
\begin{equation}\label{eq:26}
l_{k+1}-l_{k}+\frac{1}{2}e^{(1-\alpha)l_{k}}=O((\frac{1}{(k-k_{0})^{2}+\frac{2}{1-\alpha}(1/r_{k_{0}}^{2})^{1-\alpha}})^{2}).
\end{equation}
which we expect to equal zero. This means that for $k\geq k_{0}$, we expect Eqn.~\ref{eq:25} to improve
as a solution to Eqn.~\ref{eq:23} as $k_{0}$ increases (and so $r_{k_{0}}$
decreases).

\begin{figure}
\includegraphics[scale=0.4]{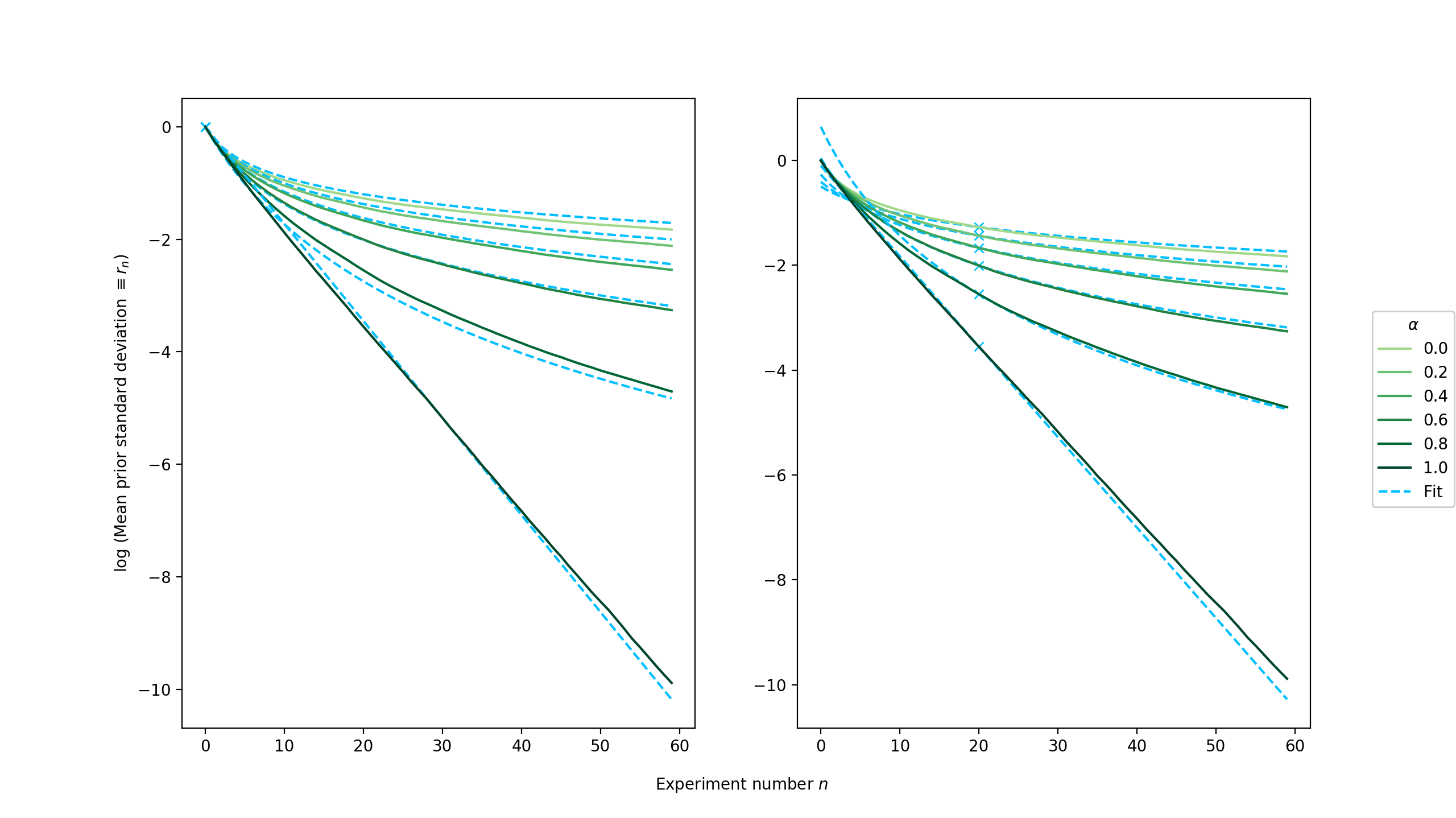}
\centering{}\caption{{\scriptsize{}Analytical solution Eqn.~\ref{eq:25} (dashed) agrees well with numerical simulations (solid) of
RFPE for different values of $\alpha$. Each simulation was performed
with 200 randomised values of the true eigenphase $\phi$ (over which
the mean is taken) and 600 samples from the posterior at each iteration
obtained by rejection filtering. The plots on the left and right figures
use initial conditions $(k_{0},r_{k_{0}})=(0,r_{0}\protect\coloneqq1)$
and $(20,r_{20})$ respectively. The fit through $(20,r_{20})$ is
more accurate for $k\geq k_{0}$ - this is expected because $r_{k}$
decreases as $k$ increases, which improves all approximations based
on $r_{k}$ small.}}
\label{Analytical numerical fit} 
\end{figure}

\begin{figure}
\includegraphics[scale=0.4]{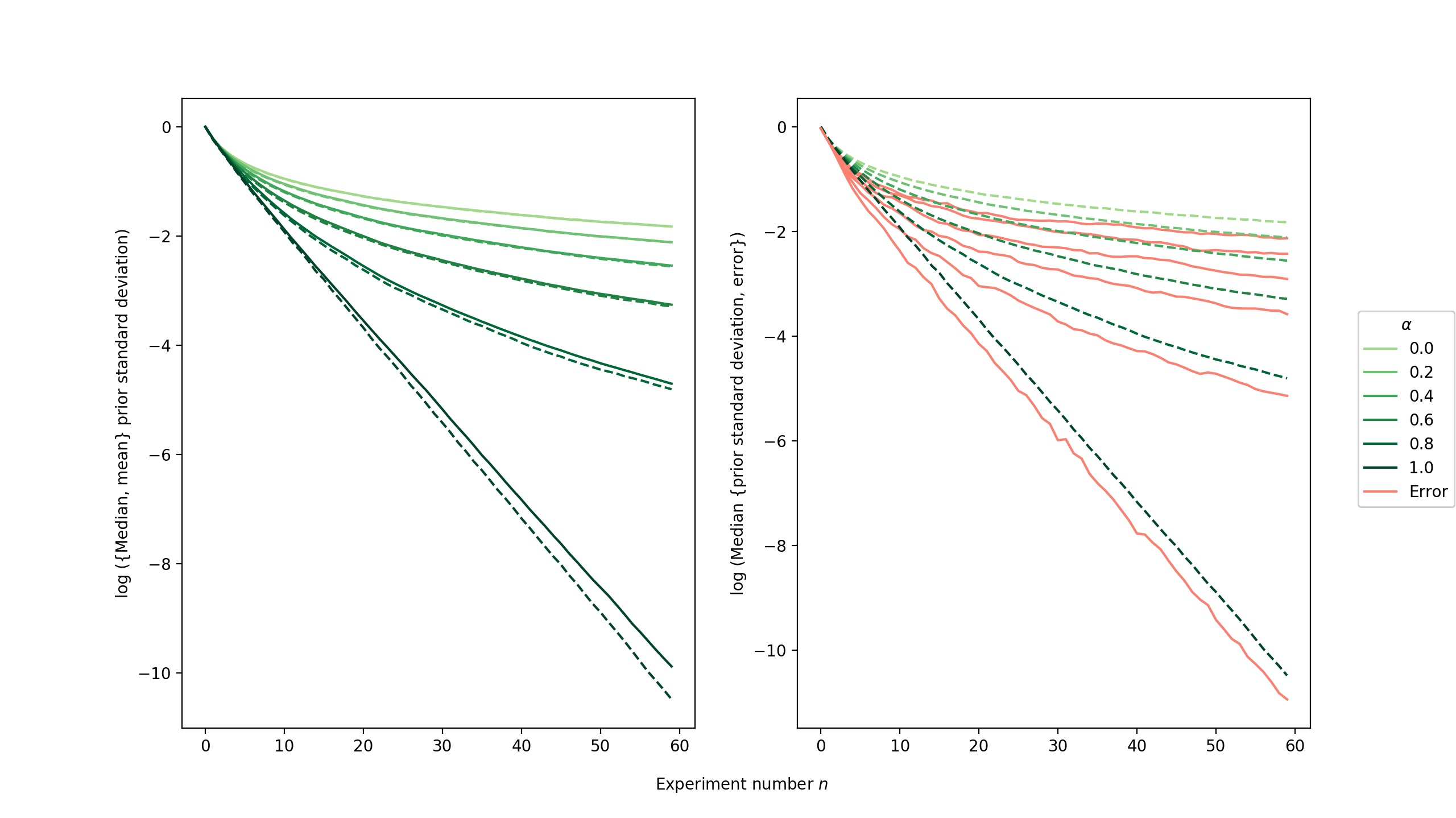}
\centering{}
\caption{{\scriptsize{} Left: We find good agreement between the analytical mean standard
deviation of Eqn.~\ref{eq:25} (dashed) and numerical median standard deviation (solid). Right: Eqn.~\ref{eq:25} (dashed) agrees
qualitatively but not quantitatively with the median error (pink). That the median errors appear to tend toward zero would
be a consequence of the \textit{weak} asymptotic consistency of phase estimates with $k$.
This fact does not preclude the mean errors (not plotted) not tending
towards zero and in fact they do not.}}
\label{Std vs error}
\end{figure}

Given the considerable number of assumptions and approximations used to
reach an analytical expression for the Bayes risk in Eqn.~\ref{eq:25}, one
is justifiably cautious about its validity. For assurance, we plotted
Eqn.~\ref{eq:25} and Eqn.~\ref{eq:17} (the latter for completeness but with
$L^{2}$ reset to $L^{2}\approx0.708$ corresponding to $a=1$)
against numerical simulations of RFPE between iterations $0$ to $60$
with two initial conditions $(k_{0},r_{k_{0}})=(0,r_{0}\coloneqq1)$
and $(20,r_{20})$. The numerical simulations are displayed in Fig.~\ref{Analytical numerical fit} and show good agreement with our analytical Eqn.~\ref{eq:25}
and Eqn.~\ref{eq:17}. Note that Eqn.~\ref{eq:25} reduces to the form
of Eqn.~\ref{eq:17} in the $\alpha=1$ limit but not exactly because of
the inaccuracy of approximation Eqn.~\ref{eq:20} when $\alpha=1$. It is also essential to point out now that the Bayes risk is a measure of precision and not a priori a measure of accuracy (i.e. error). However, in Fig.~\ref{Std vs error}, we numerically demonstrate that the median error aligns reasonably with the mean and median Bayes risk. 

Having numerically addressed two potential caveats to  Eqn.~\ref{eq:25} in Fig.~\ref{Analytical numerical fit} and Fig.~\ref{Std vs error}, we also observe from these Figures that Eqn.~\ref{eq:25} is approximately valid for $(k_{0},r_{k_{0}})=(0,1)$. Assuming this validity, we rearrange Eqn.~\ref{eq:25} to give:
\begin{align}\label{al:2}
k & =f(r_{k},\alpha),
\end{align}
where recall $f:\mathbb{R}\times[0,1]\rightarrow\mathbb{R}$ is the continuous function:
\begin{equation}\label{eq:27}
f(r,\alpha)=\begin{cases}
\frac{2}{1-\alpha}(\frac{1}{r^{2(1-\alpha)}}-1) & \text{if }\alpha\in[0,1)\\[5pt]
4\,\text{log}(\frac{1}{r}) & \text{if }\alpha=1
\end{cases}.
\end{equation}
 And Eqn.~\ref{eq:18} gives:
\begin{equation}\label{eq:28}
D_{k}\coloneqq\underset{\text{\{\ensuremath{\leq}\ {\it k }iterations\}}}{\text{max}}(M)=\frac{1}{r_{k}^{\alpha}},
\end{equation}
which together give our main interpolation result
upon replacing $(k,D_{k},r_{k})$ by $(N,D,\epsilon)$. 

The replacement of $D_{k}$ by $D$ assumes we can readily prepare the eigenstate $\Ket{\phi}$ both initially and after each measurement. We have already described why this assumption is valid in the main text. \FloatBarrier
\clearpage
\section{RFPE-with-restarts}\label{subsec:RFPE-with-restarts}

Suppose we require a precision within $\epsilon\in(0,1)$, with the constraint that $(1<)D\leq D_{\text{max}}$ for some constant $D_{\text{max}}$, but that we wish to minimise $N$. Here we calculate $N$ required by RFPE-with-restarts, assuming decoherence is detected immediately at which point RFPE switches from phase estimation to statistical sampling.

Now, $1<1/r_{k}\leq D_{\text{max}}$ gives a maximum of $N_{0}=4\log(D_{\text{max}})$
iterations in this phase estimation regime. For $k>N_{0}$, RFPE-with-restarts
switches to statistical sampling with $M$ held constant at $D_{\text{max}}$. Eqn.~\ref{al:2} then gives (under change of variable $r_{k}\leftrightarrow D_{\text{max}}\, r_{k}$
throughout the derivation) the minimum number of total iterations
of RFPE-with-restarts as: 
\begin{equation}\label{eq:30}
N'_{\text{min}}=\begin{cases}
2((\frac{1}{\epsilon D_{\text{max}}})^{2}-1)+4\log(D_{\text{max}}) & \text{if }D_{\text{max}}<\frac{1}{\epsilon}\\[5pt]
4\,\text{log}(\frac{1}{\epsilon}) & \text{if }D_{\text{max}}\geq\frac{1}{\epsilon}
\end{cases}.
\end{equation}
Again, we see an inverse quadratic scaling with $D_{\text{max}}$ in the first
case. 

In fact, we find RFPE-with-restarts is always advantageous over $\alpha$-QPE (with respect to minimising Bayes risk). This can be phrased as:
\begin{align}
    N_{\text{min}}'&\leq N_{\text{min}}\\
    \text{with equality iff }& D_{\text{max}}\in[1/\epsilon,\infty),
\end{align} 
where we recall $N_{\text{min}}$ from Eqn.~7 of the main text:
\begin{equation}
N_{\text{min}} = \begin{cases}
\frac{2}{1-\text{log}(D_{\text{max}})/\text{log}(1/\epsilon)}((\frac{1}{\epsilon D_{\text{max}}})^{2}-1) & \text{if }D_{\text{max}}<\frac{1}{\epsilon}\\[5pt]
4\,\text{log}(\frac{1}{\epsilon}) & \text{if }D_{\text{max}}\geq\frac{1}{\epsilon}
\end{cases}.
\end{equation}

One way of seeing RFPE's advantage is by writing $D_{\text{max}}=1/\epsilon^{\beta}$ where $\beta\in(0,1)$
when $1< D_{\text{max}}<1/\epsilon$, giving:

\begin{align}\label{al:3}
\frac{N_{\text{min}}'}{N_{\text{min}}} & =1-\beta-\beta(1-y)\frac{\text{log}(1-y)}{y}\nonumber \\
 & =1-\beta+\beta(1-y)\sum_{j=1}^{\infty}\frac{1}{j}y^{j-1}\nonumber \\
 & <1,
\end{align}
where $y\coloneqq1-\epsilon^{2(1-\beta)}\in(0,1)$. 

Note that the $\beta$ we introduced here can be
seen as a control parameter analogous to the $\alpha$ in $\alpha$-QPE,
and RFPE-with-restarts can be reasonably called $\beta$-QPE. By the above, we immediately deduce that $\beta$-QPE also satisfies Proposition~1 with $\alpha$ replaced by $\beta$.

While $N_{\text{min}}'\leq N_{\text{min}}$, exploratory simulations
show that $\alpha\text{-}$QPE can yield better mean
accuracy (as opposed to Bayes risk which relates to mean precision) than $\beta$-QPE for a given number of iterations
and constant $D_{\text{max}}$. In any case, should $\beta$-QPE outperform $\alpha$-QPE according to a desired metric, then we can use $\beta$-VQE (obvious definition). 

\section{$\delta$ bound and state collapse}\label{subsec:State Collapse} 

Here we present a simple 2-stage method that removes the $\delta$ bound assumption on the absolute value of $A\coloneqq \Bra{\psi}U\Ket{\psi}$ and detail state collapse into $\Ket{\pm \phi}$ within this 2-stage method.

In Stage 1, we see if $|A|$ can be bounded away from $0$ and $1$ by statistical sampling $A$ a constant number of times, which also automatically gives the sign of $A$. In Stage 2, if the bound is satisfied, we continue with $\alpha$-QPE to estimate $|A|$, gaining the efficiency boost over statistical sampling; if not, we continue with statistical sampling to estimate the expectation. 

We now present an explicit minimal specialisation of the above procedure, followed by a brief comment on how to obtain more general versions - details are omitted for brevity. 

\textbf{Stage 1.} We see if we can bound $|A|$ in the interval $I \coloneqq [\cos(5\pi/12), \ \cos(\pi/12)]$ with high confidence. We do this by estimating $A$ by statistical sampling a constant number of times. Suppose our estimate of $A$ using $n$ samples is $\hat{A}$, then Hoeffding's inequality gives:
\begin{equation}
\mathbb{P}(|A-\hat{A}| \geq t) \leq 2 \exp(-\frac{1}{2}nt^2).
\label{eq: hoeffding}
\end{equation}
Explicitly, setting $n=1000, \  t=0.1$ in Eqn.~\ref{eq: hoeffding}, we find that if our estimate $\hat{A}$ has $|\hat{A}|\in \hat{I}\coloneqq[0.36,0.85]$ then:
\begin{equation}\label{eq: exp_bound}
    \mathbb{P}(|A|\in I) \geq 0.99. 
\end{equation}
If $|\hat{A}| \in \hat{I}$ we say Stage I is successful. We get the sign of $A$ for free when Stage I is successful: the probability of inferring the correct sign is larger than $0.99$ and almost $1$.

\textbf{Stage 2.} If Stage I is unsuccessful, we continue statistically sampling $A$. If Stage I is successful, we first perform state collapse by running the RFPE circuit (main text Fig.~1) twice with the choices:
\begin{align}
\begin{split}
(M_1, \theta_1) &= (2, 0 \ \ \ \ \ \ ), \\
(M_2, \theta_2) &= (1,   b_2 \pi/2),
\end{split}
\end{align}
where $b_2\in \{0,1\}$ is the result of the first measurement.

Elementary analysis following Ref.~\cite{Dobsicek2007} gives Table~\ref{m2_specialisation}. Since $|A|\in I $, we have that $\phi \coloneqq {2\arccos{(|A|)}}\in[\pi/6,5\pi/6]$. Therefore $\sin^2(\phi) \in [0.25,1]$, $(1+\sin(\phi))/2\in[0.75,1]$ and $(1-\sin(\phi))/2\in[0,0.25]$. Hence with probability at least $0.25$ we collapse into a state that has probability of either $\Ket{\phi}$ or $\Ket{-\phi}$ greater than $0.75$. On this collapsed state we can then perform $\alpha$-QPE as prescribed in the main text. During simulations, we have found that it is more effective to modify the likelihood function of Eqn.~4 in the main text to reflect the fact that the input collapsed state has small components of either $\Ket{\phi}$ or $\Ket{-\phi}$.

This concludes our explicit description of a minimal specialisation of the 2-stage method. There are many possible modifications. In particular, we may want to expand the interval $\hat{I}$ so that we are more likely to be successful in Stage 1. To do this, we can either increase the number of statistical samples we take of $A$ or more importantly, we can increase the number $m$ of measurements in Stage 2. Increasing $m$ increases our ability to resolve between $\Ket{\phi}$ and $\Ket{-\phi}$, necessary because $\phi$ can be closer to $-\phi$ when $\hat{I}$ is expanded.

\begin{table}[H]
    \centering
    \begin{tabular}[b]{c|c|c}
        Measure $(b_2,b_1)$ & Probability & Probability of $\Ket{\phi}$ \tabularnewline \hline 
        $(0,0)$ & $\cos^2(\phi)\cos^2(\phi/2)$ & $1/2$ \tabularnewline 
        $(0,1)$ & $\cos^2(\phi)\sin^2(\phi/2)$ & $1/2$ \tabularnewline
        $(1,0)$ & $\sin^2(\phi)/2$ & 
        $(1+\sin{\phi})/2$
        \tabularnewline
        $(1,1)$ & $\sin^2(\phi)/2$ & $(1-\sin{\phi})/2$
    \end{tabular}
    \caption{Measurement probabilities and the probability of $\ket{\phi}$ in the collapsed $\Ket{\psi}$ given the 4 possible measurement outcomes when performing $m=2$ measurements. Expressions for when performing $m>2$ measurements are also straightforward to derive but are omitted for brevity.} 
    \label{m2_specialisation}
\end{table}

\FloatBarrier
\end{document}